\documentstyle[prl,aps]{revtex}
\begin{document}
\title
{Why General Theory of Relativity Allows the Existence of Only Extremal
Black Holes}

\author{Abhas Mitra}
\address{Theoretical Physics Division, Bhabha Atomic Research Center,\\
Mumbai-400085, India\\ E-mail: amitra@apsara.barc.ernet.in}


\maketitle

\begin{abstract}
Supersymmetric String theories find occurrences of extremal Black
Holes with gravitational mass $M=Q$ where $Q$ is the charge
($G=c=1$). Thus, for the chargeless cases, they predict $M=0$. We
show that General Theory of Relativity, too, demands a unique BH
mass $M=0$. Physically, this means that, continued gravitational
collapse indeed continues for {\em infinite proper time} as the
system hopelessly tries to radiate its entire original mass
energy to attain the lowest energy state $M=0$.
\end{abstract}

\vskip 0.5 cm
The concept of Black Holes (BHs) is now important not only for
astrophysicists but also for elementary particle physicists. In
particular, one of the promising candidates for the Quantum
Gravity is the Supersymmetric String theory (or M-theory). In the
low energy limit, such theories are naturally expected to be
consistent with classical General Theory of Relativity (GTR).
However, the supersymmetric theories find the occurrences of BHs
with mass $M=0$ for the chargeless Schwarzschild case\cite{1}. We
show that, GTR too, actually, yields the same result. In the
context of the GTR, the concept of BHs first arose with the
discovery of famous {\em vacuum} spherically symmetric
Schwarzschild solution\cite{2}:
\begin{equation}
ds^2 = \left( 1 - {r_g \over r}\right) dt^2 -  (1 -{r_g\over
r})^{-1} dr^2
- r^2 (d\theta^2 + \sin^2 \theta d\phi^2)
\end{equation}
Here we shall call `$r$'' as the ``Schwarzschild radial
coordinate''; $\theta$ and $\phi$ are the usual polar angles and
$r_g =2M$ ($G=c=1$). Because of the importance of this solution,
we would like to remind the reader the salient points behind it
as discussed by Landau \& Lifshitz\cite{2}. Bearing in mind the fact that
there can not be any spacetime cross term in the metric
describing isotropic cases, the most general form of the metric (not
necessarily for vacuum)
is
\begin{equation}
ds^2 = e^{\nu(r)} dt^2 - e^{\lambda(r)} dr^2 - e^{\mu(r)} (d\theta^2
+\sin^2\theta d\phi^2)
\end{equation}
But for a spherically symmetric {\em finite system}, the origin is unique
and out of infinite
choices for the value of $r$, only one value is physically meaningful.
 And this uniqueness comes from the function $e^\mu(r)$. For a
fixed $r$ and $t=constant$ hypersurface, and for a fixed
$\theta=0$, the invariant circumference is
\begin{equation}
\oint ds = 2\pi   e^{\mu/2}
\end{equation}
And therefore only the {\em unique} and invariant choice
$e^{\mu/2} =r$
 can claim to be the
physically meaningful radial coordinate.
 And thus
$r$ naturally retains its essential space like character in every
situation. By starting from this general premises, one can derive
the {\em vacuum} Schwarzschild metric (1) without imposing a single assumption
or  extraneous condition like whether $ r >2M$ or $r \le 2M$.
 One important point is that this metric
is,  naturally, asymptotically flat, as can be seen that
$g_{rr} =g_{tt} =1$ at $r=\infty$. And then, by demanding that at
large $r$, the metric assumes Newtonian form, one interprets $M$,
appearing in Eq. (1) as the ``gravitational'' mass.
Simultaneously $t$, as defined by Eq. (1),  assumes a clear
physical meaning as the {\em proper time} of $S_\infty$, the
distant inertial observer. And since, there is {\em no assumption
or precondition} in the derivation of Eq. (1), naturally,
 Schwarzschild
 was unhappy with the occurrence of this singularity in his
solution and consoled himself with the fact that no physical body, in
hydrostatic equilibrium, can be squeezed below $r <2M$. If a body is
static and could be squeezed at $r=2M$, its surface gravitational
red-shift would be $z=\infty$:
\begin{equation}
z_s =(1-2M/r)^{1/2} -1
\end{equation}
On the other hand, Schwarzschild found that the absolute upper limit $z_s
\le 2$ for any static body.

Here recall that for a free falling particle the radial speed as
measured by a local static Schwarzschild observer $V_{Sch} =1$ at
$r =2M$. Then no amount of Lorentz boost can hold the Sch.
observer as {\em static} and this fact is turned around to say
that the Sch. coordinate breaks down at $r=2M$. Physically this
would mean ``the system of reference for $r <r_g$ can be realized
only by means of moving bodies, whose motion is directed toward
the center''\cite{2}. In practical terms this would mean that, in
case we are trying to model a relativistic {\em static star}
having $r
\le 2M$, we would not be successful, and on the other hand, the
star must be collapsing, which in turn means that, when one is
describing the collapse of a dust ball, he is free to match the
internal solutions with the external vacuum Sch. solutions. And
indeed, Oppenheimer and Snyder (OS)\cite{3} did so in their
famous work. Nonetheless,
 such explanations hardly appear to be  satisfactory because if
 $r$ has a clear physical significance as the {\em invariant radius} and $t$
too has a clear physical significance as the {\em time recorded}
by $S_\infty$, why should the Sch. coordinate break down at
$r=2M$ (for the static case) or why should the metric
coefficients be singular at $r=2M$? Rather than ever trying to
face such physical questions head on, traditionally all authors,
 have, inadvertently,
pushed them below the carpet of mathematics, by using the
standard refrain that it is like the singularity at the origin of
the polar coordinate system ($g_{\phi \phi}
=0$ at $\theta$ =0).  And the freedom of choice of coordinates in GTR has
come very handy in running away from such poignant physical
questions. For example, it was generally agreed upon by the
community in 1960 that the Kruskal coordinates\cite{4}, rather than the $r, t$
coordinates correctly describes the spacetime both inside and
outside of a Schwarzschild BH. For the external region (Sector I)
\begin{equation}
u=f_1(r) \cosh
{t\over 4M}; \qquad v=f_1(r) \sinh
{t\over 4M}; r\ge 2M
\end{equation}
where
\begin{equation}
f_1(r) = \left({r\over 2M} -1\right)^{1/2} e^{r/4M}
\end{equation}
However, if one would stick to this definition of $u$ and $v$,
they would be imaginary for $r <2M$. But since $u$ and $v$ are
believed to be the real physical coordinates (rather than $r,
t$), {\em they can not be allowed to be imaginary}. Therefore, by hand,
the definition of $u$ and $v$ are altered
 for the region interior to the {\em supposed} event horizon (Sector II):
\begin{equation}
u=f_2(r) \sinh
{t\over 4M};\qquad v=f_2(r) \cosh
{t\over 4M}; r\le 2M
\end{equation}
where
\begin{equation}
f_2(r) = \left(1- {r\over 2M}\right)^{1/2} e^{r/4M} =\sqrt{-1} f_1(r)
\end{equation}
But note that, even now, {\bf to ensure that $u$ and $v$ are
definable at all, first of all $r$ and $t$ must be definable over
the entire spacetime}. But how can $r$ and $t$ be meaningfully
defined for $r< 2M$ if either of them ceases to be definable?
Note that, if we stick to the original relationship between $r$
and $t$ (like that of OS) $t$ would become {\em undefinable} or
imaginary for $r<2M$
\begin{equation}
{t\over 2M} = \ln{(r_\infty/2M-1)^{1/2} + \tan{(\eta/2)} \over (r_\infty
/2M-1)^{1/2} - \tan{(\eta/2)}} + \left({r_\infty\over 2M}-1\right)^{1/2}
\left[\eta + \left({r_\infty\over 4M}\right)(\eta +\sin \eta)\right]
\end{equation}
Here the test particle is assumed to be at rest at $r=r_\infty$
at $t=0$ (or for dust collapse, the starting point) and the
``cyclic coordinate'' $\eta$ is defined by
\begin{equation}
r= {r_\infty\over 2} (1 + \cos \eta)
\end{equation}
Since $\tan(\eta/2) = (r_\infty/r -1)^{1/2}$, we may rewrite Eq. (9)
in terms of a new variable
\begin{equation}
x = \left({r_\infty /2M -1\over r_\infty /r_b -1}\right)^{1/2}
\end{equation}
as
\begin{equation}
{t\over 2M} = \ln  {x +1\over
x-1} + \left({r_\infty \over 2M}
-1\right) \left[\eta + \left({r_\infty\over 4M}\right) (\eta + \sin \eta)\right]
\end{equation}
It can be easily found that, irrespective of $M$ being finite or zero, $t
\rightarrow \infty$ as $x \rightarrow 1$ or $r \rightarrow 2M$.
For the time being, we assume that $M \ge0$. Then, from
Eq. (11), note that
\begin{equation}
x \le 1 ; \qquad for~ r\le 2M
\end{equation}
Thus for $r <2M$,
 $t$ is {\em not definable at all} because the argument of
logarithmic function can not be negative. However nobody seems to have pondered
 {\em whether the
situation which is leading to an imaginary $t$ is unphysical or
not}. How can $t$ be imaginary if $r$ remains real with its glory
as the physically measurable ``invariant radius''. Note, even
when one purports to describe the EH or the central singularity,
one does so in terms of $r$, i.e, whether $r =2M$ or whether
$r=0$. And, as far as $S_\infty$ is concerned, he is either able
to watch an event ($t=finite$) or unable to do so ($t=\infty$).
Probably later authors\cite{5} realized that $t$ can not be
allowed to be imaginary, not because of the fact that $t$ is
still the {\em proper time} of a Galilean observer, a measurable
quantity, but because of the fact that, otherwise esoteric new
coordinates, like $u$ and $v$ would not be definable. Thus we
find that a modulus sign was introduced in the $t-r$
relationship\cite{5}:
\begin{equation}
{t\over 2M} = \ln \mid {x +1\over
x-1}\mid + \left({r_\infty \over 2M}
-1\right) \left[\eta + \left({r_\infty\over 4M}\right) (\eta + \sin \eta)\right]
\end{equation}
But, the conceptual catastrophe actually becomes worse by this
tailoring. Of course, as the particle enters the EH, still, $t
\rightarrow \infty$. Note that, $t$ having become infinite,
definitely can not decrease, and more importantly {\bf inside the
EH, $t$ can not be finite} because, otherwise, {\em it would
appear that although the distant observer can not witness the
exact formation of the EH (in a finite time), nevertheless, he
can witness the collapse of the fluid inside the EH}. This would
mean violation of causality and the existence of some sort of a
``time machine''. And we know that, it is only the comoving
observer who is supposed to witness both the formation of EH and
the collapse beyond it. But the foregoing equation tells that
 if $M >0$ (finite), the {\em the value of $t$ not
only starts decreasing} but also suddenly {\bf becomes finite} as
the boundary enters the EH ($r <2M$)!! And as the collapse is
complete,
\begin{equation}
x =0; \qquad ~if~ M >0; \qquad at~ r=0
\end{equation}
 the corresponding value of $t=T$ required by the distant observer to
see the {\bf collapse to the central singularity within the EH} is simply
\begin{equation}
t = T = 2M
 \left({r_\infty\over 2M}
-1\right) \left[\eta + \left({r_\infty\over 4M}\right) (\eta + \sin \eta)\right]
\end{equation}
By using Eq. (5), this can be rewritten as
\begin{equation}
T= \pi (r_\infty - 2M) \left( 1+ {r_\infty\over 4M}\right)
\end{equation}
And note that, if we really insist that $T=\infty$ as per the
original agenda, we must realize that $M=0$! This would mean (i)
there is no additional spacetime between the EH and the central
singularity, i.e, they are synonymous and the Sch. singularity is
a genuine singularity provided one realizes that by the time one
would have $r=2M$, the value of $M\rightarrow 0$, and (ii) the
proper time for formation of this singularity is $\tau \propto
M^{-1/2} =\infty$. The latter means that, it is not formed at any
finite proper time, the collapse process continues indefinitely
and there is {\em no incompleteness for the timelike geodesics}.

 Einstein, too, was equally worried about
 this singularity and his intuition (correctly) told that it can not occur
in actual physical cases. In fact he (unsuccessfully) struggled
in 1938 to show that it was indeed so\cite{6}. But his proof was
not convincing and was ignored. On the other hand, in 1939, it
was convincingly shown by Oppenheimer and Volkoff\cite{7}, that
given a certain equation of state (EOS) there is an upper limit
on the gravitational mass upto which it is possible to have
static configurations. Coupled with the fact that $z_{s} \le
2$\cite{1}, it definitely meant that sufficiently massive bodies
will undergo continued gravitational collapse. But the most
important problem left to be answered now was where does the
collapse process stop, a question actually burning ever since
Newton discovered gravitation as an universal property. In the
same year 1939, OS\cite{3} set out to find an answer for this
question by solving the Einstein equations for an idealized
homogeneous ``dust''. They attempted to explicitly find the
asymptotic behaviour of the metric coefficients for $r \le 2M$ as
a function of $t$ because, as explained earlier, {\em for a
collapsing body}, $r,t$ remains valid coordinate system even when
one (incorrectly) considers $M >0$. For the internal solutions,
for the sake of convenience, they first worked with ``comoving
coordinates'' $R$ and $\tau$ and then transformed the results to
$r,t$ system by matching the solution with Eq. (1) at the
boundary. And their solution (apparently) gave the impression
that gravitational collapse ends in the formation of a BH of
unspecified gravitational mass $M$. Since then, it has not been
possible to find {\em any other exact solution of gravitational
collapse even for the spherically symmetric case}. And inspired
by the OS work, in the sixties and seventies a large number of
relativists, formulated, by trials and errors, the so called
``singularity theorems''\cite{8} which showed that, under a set
of (apparently) reasonable assumptions, generic gravitational
collapse should result in the formation of ``singularities''.
These singularities could be both ``naked'' as well as BH type.
And the Cosmic Censorship Conjecture\cite{9} asserts that the
resultant singularity should be a BH type. One of the important
assumptions behind the singularity theorems is that of the
existence of ``trapped surfaces'', which for the spherical case
implies the occurrence of a surface with
\begin{equation}
{2M(r) \over r} > 1
\end{equation}
However in 1998, it was shown by us that independent of the details of the
collapse (or expansion) like the EOS and radiation transport properties,
{\em trapped surfaces do not form}\cite{10} :
\begin{equation}
{2M(r)\over r} \le 1
\end{equation}
 Consequently, for continued collapse, the final gravitational
mass
\begin{equation}
M(r) \rightarrow 0; \qquad r\rightarrow 0
\end{equation}
if we rule out occurrence of negative mass (repulsive gravity).
Physically, $M=0$ state corresponds to the {\em lowest energy
state} which, the system naturally strives to attain by radiating
the entire original mass energy. We have also found that the
physically meaningful collapse ends with $2M/r =1$ state rather
than $2M/r <1$ state\cite{11}. Thus as if the system tries to
attain a zero mass BH state. Eventually, it is found that the
proper time for formation of this state is $\tau
=\infty$, which implies that GTR is singularity free (atleast for
isolated bodies). Since our results are most general\cite{10}, it
is trivially applicable to the OS case too. And naturally this
result must be imprinted in the explicit OS solutions.
Unfortunately, probably because of the excitement of having found
BH solutions, OS\cite{3} completely overlooked this telltale
imprint in their Eq. (36). They instead attempted to simplify Eq.
(36) in such a manner that this explicit imprint got obliterated.
Consequently they obtained physically inconsistent solutions
under the assumption of a finite value of $M$. In fact they
partially admitted the inconsistent internal solutions: for an
internal point ($R <R_b$), $e^\lambda$ {\em refused to become
infinite even at} $r=0$, i.e, when the collapse was complete!
Unfortunately, they did not care to pursue this matter further
(may be because of 2nd world war) and sent their paper to
Physical Review where it got accepted. And the rest is,
 of course, history.
 And it is only recently that we have brought out all such
issues\cite{12} in a most transparent manner to show that the OS
work itself demanded $M=0$ even in the absence of our general
result (Eq. [19]).

Yet to be doubly sure about the non-existence of finite mass BHs, in
another recent work\cite{13}, we first assumed the existence of a finite
mass Schwarzschild BH.
The radial part of the Kruskal metric has the form
\begin{equation}
ds^2 = g_{vv} dv^2 - g_{uu} du^2
\end{equation}
On the positive side all that Kruskal transformations achieved
was to ensure that $g_{uu} = -g_{vv}$ was definable over the
entire spacetime (provided of course $r$ and $t$ are definable in
the first place), they retained their respective original
algebric signs, and  do not blow up at $r=2M$
(provided $M >0$). On the flip side, it created a pandora's box
as far as physical concepts are concerned. For instance, it
demanded that (i) although the central singularity is still
described by $r=0$, its actual structure is that of a pair of
hyperbolas : $u = \pm (1+v^2)^{1`/2}$, (ii) the negative branch
of the hyperbola corresponds to ``White Hole'' which can spew out
mass energy at its will, (iii) Inside the BH, there are two
universes connected by a spacetime ``throat'', (iv) Even the
spacetime at $r =\infty$, which is naturally seen to be flat and
Newtonian by the original and exact Sch. solution, has a complex
structure corresponding to $u \ge \pm \mid v\mid$ or $u \le \pm \mid v\mid$\cite{4,5}.

To the knowledge of the present author, nobody raised here the
question, when there are expected to be $N$ BHs, how much complex
will be structure of the spacetime far away from all the BHs? And
when astronomical observations firmly supported the view that far
from massive objects, the structure of the spacetime is well
described by the mundane $r,t$ coordinates, there was no
introspection as to whether
 the complex structure of spacetime
at $r=\infty$, suggested
 by the Kruskal view, was acceptable or not.
 Note,
if a finite mass BH were a physical object, the {\em
radial geodesic of a physical particle must remain timelike at EH}. And we
have directly shown that it is not so; it becomes {\em null just like the
Schwarzschild case}! Any reader can verify this by noting that since as
$r\rightarrow 2M$, $u \rightarrow \pm v$, and consequently the Kruskal
derivative assumes a form\cite{13}
\begin{equation}
{du\over dv} \rightarrow {f(r,t, dt/dr)\over \pm f(r,t, dt/dr)}; \qquad
r\rightarrow 2M
\end{equation}
And the corresponding limit becomes unity irrespective of whether
$f\rightarrow 0, \infty$ or anything. Therefore, one has $du^2
= dv^2$ at $r=2M$ in Eq. (21), so that $ds^2 =0$ !

Physically this means that the free fall speed at the EH, $V =1$,
and this is not allowed by GTR unless $R=M=0$. Now we explain why
$V =1$ at the EH for any coordinate system, Kruskal or Lemaitre
or anything else. Let the speed of the static other observer be
$V_{Sch-O}$ with respect to the Schwarzschild observer. By
principle of equivalence, we can invoke special theory of
relativity locally. Then the free fall speed of the material
particle with respect to the other static observer will be
\begin{equation}
V = {V_{Sch} \pm V_{Sch -O} \over 1 \pm V_{Sch} V_{Sch-O}}
\end{equation}
But since $V_{Sch} =1$ at the EH, we would always obtain $\mid
V\mid=1$ too. {\bf And hence there can not be any finite mass
BH}. The value of $V$ can change in various coordinates only as
long as it is subluminous to all observers. Thus, all that
Kruskal transformations and several other transformations tried
to do was to arrive at a metric whose coefficients do not (appear
to) blow up at $r=2M$. But as we saw, this was a purely cosmetic
approach because no independent effort was ever made to verify
whether the actual value of $ds^2 =dt^2 (1- 2M/r)$, which becomes
zero at $r=2M$ for a Schwarzschild observer becomes timelike,
$ds^2 >0$, in the new coordinate.

{\bf And since after all, $ds^2$ is an invariant, the value of it
can not change} unless, in the new coordinate, the location of
the EH materially changes! The latter statement means that, the
location of the Sch. singularity or the central singularity have
to be described independent of the clutches of the Schwarzschild
system. In other words, the new coordinates must not be obtained
by transforming the $r, t$ coordinates, if it were possible. It
is really inexplicable how almost all the authors overlooked this
simple point while taking the existence of finite mass BHs for
granted. We still hope that the reader will not reject
this paper to uphold the same inexplicable legacy. Finally, we
conclude that as far as the value of the mass of the BHs are
concerned, the results obtained by Supersymmetric String theories
completely agree with corresponding GTR results. This result is
also, in a certain way, in agreement with the result that the
naked singularities could be of zero gravitational mass\cite{14}.

\end{document}